\documentclass{PoS}

\usepackage{latexsym,amssymb,amsmath,amsbsy,epsfig,multirow}

\usepackage{stmaryrd,xcolor}
\usepackage{amsfonts,graphicx,float,ulem}
\usepackage{bbm}
\usepackage{color,graphicx}

\newcommand{\be}{\begin{equation}}
\newcommand{\ee}{\end{equation}}
\newcommand{\ba}{\begin{eqnarray}}
\newcommand{\ea}{\end{eqnarray}}
\newcommand{\bs}{\begin{subequations}}
\newcommand{\es}{\end{subequations}}
\newcommand{\no}{\nonumber\\}

\allowdisplaybreaks

\title{Higgs masses and couplings in the general 2HDM with unitarity bounds}

\ShortTitle{Higgs masses and couplings in the general 2HDM with unitarity bounds}

\author{\speaker{D.~Jur\v{c}iukonis}\thanks{D.J. thanks the
Lithuanian Academy of Sciences for support through the project DaFi2018.}$\phantom{.}^{(1)}$ and L.~Lavoura$^{(2)}$ \\
		$^{(1)} \! $		
		Vilnius University,
		Institute of Theoretical Physics and Astronomy \\
		$^{(2)} \! $	        
        CFTP, Instituto Superior T\'ecnico, Universidade de Lisboa \\        		
        E-mail: \email{darius.jurciukonis@tfai.vu.lt}}

\abstract{We investigate the general two Higgs doublet model
  imposing both the unitarity conditions
  and the bounded-from-below conditions.
  Both types of conditions restrict the ranges of the parameters
  of the scalar potential.
  We study the model in the Higgs basis,
  \textit{i.e.}\ in the basis for the scalar doublets
  where only one doublet has vacuum expectation value.
  We use the experimental bounds on the oblique parameter $T$,
  to produce scalar particles with masses
  and cubic and quartic couplings of the Higgs
  in agreement with the phenomenology.
  The numerical calculations show that
  the cubic coupling may be up to 1.6 times larger than
  in the Standard Model,
  but it may also be zero or even negative.
  The quartic coupling is always positive
  and may be up to four times larger than in the Standard Model.
}

\FullConference{The 39th International Conference on High Energy Physics (ICHEP2018)\\
		4-11 July, 2018\\
		Seoul, Korea}
		
\begin{document}

\section{The model}

The Standard Model (SM) predicts a boson $h_1$ which is a scalar
and it predicts its cubic and quartic couplings $g_3$ and $g_4$,
which we define through
\be
\label{kmhiho}
\mathcal{L} = \cdots - g_3 \left( h_1 \right)^3 - g_4 \left( h_1 \right)^4,
\ee
to be $g_3^\mathrm{SM} \approx 32$\,GeV and $g_4^\mathrm{SM} \approx 0.032$,
respectively.
But Nature could be more complicated than the SM
and then $g_3$ and $g_4$ might have different values.
In this paper we analyse the cubic $g_3$ and quartic $g_4$ couplings
of the model with two scalar gauge-$SU(2)$ doublets
$\phi_1$ and $\phi_2$ having the same weak hypercharge.
This is usually known as 2HDM.
The most general scalar potential is
\ba
\label{gjihoree}
V &=& \mu_1 \phi_1^\dagger \phi_1 + \mu_2 \phi_2^\dagger \phi_2
+ \left( \mu_3 \phi_1^\dagger \phi_2 + \mathrm{H.c.} \right) 
\hspace*{7mm} \no & &
+\frac{\lambda_1}{2} \left( \phi_1^\dagger \phi_1 \right)^2
+ \frac{\lambda_2}{2} \left( \phi_2^\dagger \phi_2 \right)^2
+ \lambda_3\, \phi_1^\dagger \phi_1\, \phi_2^\dagger \phi_2
+ \lambda_4\, \phi_1^\dagger \phi_2\, \phi_2^\dagger \phi_1
\hspace*{7mm} \no & &
+ \left[
  \frac{\lambda_5}{2} \left( \phi_1^\dagger \phi_2 \right)^2
  + \lambda_6\, \phi_1^\dagger \phi_1\, \phi_1^\dagger \phi_2
  + \lambda_7\, \phi_2^\dagger \phi_2\, \phi_1^\dagger \phi_2
  + \mathrm{H.c.}
  \right],
\ea
where $\mu_{1,2}$ and $\lambda_{1,2,3,4}$ are real.

We use the Higgs basis for the scalar doublets where only $\phi_1^0$ has VEV
\be
\label{nuiho}
\phi_1 = \left( \begin{array}{c} G^+ \\ v + \left( H + i G^0 \right)
  \left/ \sqrt{2} \right. \end{array} \right),  \hspace{1cm}
\phi_2 = \left( \begin{array}{c} C^+ \\ \left( \sigma_1 + i \sigma_2 \right)
  \left/ \sqrt{2} \right. \end{array} \right).
\ee
Here $v$ is the VEV,
which is real and positive,
and $G^+$ and $G^0$ are (unphysical) Goldstone bosons.
In the second equation,
$\sigma_1$ and $\sigma_2$ are real fields
and $C^+$ is the physical charged scalar of the 2HDM.

The unitarity constraints lead to upper bounds
on the parameters of the potential.
The idea of these constraints is that
the scalar--scalar scattering amplitudes at tree-level
must respect unitarity.
For the 2HDM there are five two-particle scattering channels
having different values of the electric charge $Q$
and of the third component of weak isospin $T_3$.
In order to derive the unitarity conditions
one must write the scattering matrices
for pairs of one incoming state and one outgoing state
with the same $Q$ and $T_3$.
In order to satisfy those conditions,
the eigenvalues of all the scattering matrices should be smaller,
in modulus, than $4 \pi$.
These conditions were first derived in ref.~\cite{unitarity}
and the expressions of matrices for our model are presented in ref.~\cite{musu}.

The potential~\eqref{gjihoree} must be positive
in all the field-space directions for large values of the fields;
this means that the  scalar potential has to be bounded from below (BFB),
which provides stability for the potential.
Necessary and sufficient conditions for the scalar potential of the 2HDM
to be BFB were derived in ref.~\cite{maniatis,silva-ivanov},
which we have implemented in our numerical calculations~\cite{musu}. 
Also we apply conditions~\cite{vacuum}, which guarantee that the vacuum state has a lower value of the potential than all the other possible stability points of the potential. %, \textit{i.e.} potential is in the global minimum. 

We emphasize that both the unitarity conditions
and the bounded-from-below conditions for the 2HDM
are invariant under a change of the basis used for the two doublets.
Therefore, one may implement those conditions directly in the Higgs basis.

The mass terms of the scalars are
\be
V = \cdots + \frac{1}{2}
\left( \begin{array}{cccc} H & \sigma_1 & \sigma_2  \end{array} \right)
M
%\left( \begin{array}{c} H \\ \sigma_1 \\ \sigma_2 \\  \end{array} \right),
\left( \begin{array}{cccc} H & \sigma_1 & \sigma_2  \end{array} \right)^T,
\ee
with mass matrix
\bs
\label{Mmatr}
\ba
\label{Mmatr1}
M &=& \left( \begin{array}{ccc}
  2 \lambda_1 v^2 & 2 v^2\, \Re{\lambda_6} & - 2 v^2\, \Im{\lambda_6} \\
  2 v^2\, \Re{\lambda_6} &
  M_C + \left( \lambda_4 + \Re{\lambda_5} \right) v^2 &
  - v^2\, \Im{\lambda_5} \\
  - 2 v^2\, \Im{\lambda_6} &
  - v^2\, \Im{\lambda_5} &
  M_C + \left( \lambda_4 - \Re{\lambda_5} \right) v^2
\end{array} \right) \hspace*{5mm}
\\*[2mm] &=& \label{MMM}
R^T \times \mathrm{diag} \left( M_1,\ M_2,\ M_3 \right) \times R,
\ea
\es
where $R$ is a $3 \times 3$ orthogonal matrix
that may be parameterized as
\be
\label{RRR}
R = \bar{R}_{23}(\vartheta_2) \times \bar{R}_{12}(-\vartheta_1) \times \bar{R}_{23}(-\vartheta_3),
\ee
%
%%
%\be
%\label{RRR}
%R = \left( \begin{array}{ccc}
%  c_1 & s_1 c_3 & s_1 s_3 \\
%  - s_1 c_2 & c_1 c_2 c_3 + s_2 s_3 & c_1 c_2 s_3 - s_2 c_3 \\
%  - s_1 s_2 & c_1 s_2 c_3 - c_2 s_3 & c_1 s_2 s_3 + c_2 c_3
%\end{array} \right).
%\ee
%%
%Here, $c_j = \cos{\vartheta_j}$ and $s_j = \sin{\vartheta_j}$ for $j = 1, 2, 3$. 
by using rotation matrices $\bar{R}_{12}$ and $\bar{R}_{23}$. 
The squared mass $M_1 = \left( 125\, \mathrm{GeV} \right)^2$ and $M_2 < M_3$.
We do not impose any lower limit on $M_{2,3}$;
we allow them to be lower than $M_1$. 

According to equation~\eqref{gjihoree} the three-Higgs vertex is given by
%
%\bs
\ba
\label{g3}
g_3 &=& \frac{v}{\sqrt{2}} \left[
  \lambda_1 x_1^3
  + \left( \lambda_3 + \lambda_4 \right) x_1 \left( 1-x_1^2 \right)
  + x_1 \left( x_2^2 - x_3^2 \right) \Re{\lambda_5}
  - 2 x_1 x_2 x_3 \, \Im{\lambda_5}
  \right. \notag \\ & & \left.
  + 3 x_1^2 \left( x_2\, \Re{\lambda_6} - x_3\, \Im{\lambda_6} \right)
  + \left( 1-x_1^2 \right) \left( x_2\, \Re{\lambda_7} - x_3\, \Im{\lambda_7} \right)
  \right],
\ea
%\es
%
and the four-Higgs vertex is given by
%
%\bs
\ba
\label{g4}
g_4 &=&
\frac{1}{8} \left[ \lambda_1 x_1^4
+ \lambda_2 \left( 1-x_1^2 \right)^2
+ 2 \left( \lambda_3 + \lambda_4 \right) x_1^2 \left( 1-x_1^2 \right)
+ 2 x_1^2 \left( x_2^2 - x_3^2 \right) \Re{\lambda_5}
\right. \notag \\ & & \left. 
- 4 x_1^2 x_2 x_3\, \Im{\lambda_5}
+ 4 x_1^3 \left( x_2\, \Re{\lambda_6} - x_3\, \Im{\lambda_6} \right)
+ 4 x_1 \left( 1-x_1^2 \right)\left( x_2\, \Re{\lambda_7} - x_3\, \Im{\lambda_7} \right)\right],
\ea
%\es
%
where $x_k = R_{1k}$ for $k = 1, 2, 3$
and $R$ is the matrix in equation~\eqref{RRR}.

%The Oblique parameter is given by
%%
%\bs
%\ba
%T &=& \frac{1}{16 \pi s_w^2 m_W^2} \left\{
%  \left( 1 - \bar{x}_1 \right) F \left( M_C,\ M_1 \right)
%  + \left( 1 - \bar{x}_2 \right) F \left( M_C,\ M_2 \right)
%  \right. \\ & &
%  + \left( 1 - \bar{x}_3 \right) F \left( M_C,\ M_3 \right)
%  - \bar{x}_1\, F \left( M_2,\ M_3 \right)
%  - \bar{x}_2\, F \left( M_1,\ M_3 \right)
%  \\ & &
%  - \bar{x}_3\, F \left( M_1,\ M_2 \right)
%  + 3 \left( 1 - x_1 \right) \left[ F \left( M_1,\ M_W \right)
%    - F \left( M_1,\ M_Z \right) \right]
%  \\ & & \left.
%  - 3\, \bar{x}_2 \left[ F \left( M_2,\ M_W \right)
%    - F \left( M_2,\ M_Z \right) \right]
%  - 3\, \bar{x}_3 \left[ F \left( M_3,\ M_W \right)
%    - F \left( M_3,\ M_Z \right) \right]
%  \right\}
%\ea
%\es
%%
%where
%%
%\be
%F \left( x,\, y \right) = \left\{
%\begin{array}{lcl}
%  {\displaystyle \frac{x + y}{2} - \frac{x y}{x - y}\, \ln{\frac{x}{y}}} &
%      \Leftarrow & x \neq y, \\
%  0 & \Leftarrow & x = y,
%\end{array} \right.
%\ee
%%
%and $\bar{x}_1 = (R_{11})^2$, $\bar{x}_2 = (R_{21})^2$ and $\bar{x}_3 = (R_{31})^2$.

\section{Numerical analysis}

The potential~\eqref{gjihoree} of the 2HDM,
in the Higgs basis,
has 11 real parameters:
$v$,
$\lambda_1$,
$\lambda_2$,
$\lambda_3$,
$\lambda_4$,
$\left| \lambda_5 \right|$,
$\left| \lambda_6 \right|$,
$\left| \lambda_7 \right|$,
$\arg{\left( \lambda_6^\ast \lambda_7 \right)}$,
$\arg{\left( \lambda_5^\ast \lambda_6 \lambda_7 \right)}$,
and $\mu_2$.
The other three real parameters of the potential depend on these:
$\mu_1 = \lambda_1 v^2$ and $\mu_3 = \lambda_6 v^2$ ($\mu_3$ is complex,
therefore it represents two real parameters).
By putting these 11 parameters into the mass matrix~\eqref{Mmatr1}
we can compute its eigenvalues $M_{2,3}$,
diagonalizing matrix $R$,
and the Higgs couplings $g_3$ and $g_4$.
A detailed description of this calculation method
is presented in ref.~\cite{musu}.
It turns out that this method prefers to produce largish values of the masses
and misses lower masses. 

In order to have control on the values of the masses,
in our numerical work we can use as input different quantities. 
Besides $v = 174$\,GeV and $M_1 = \left( 125\,\mathrm{GeV} \right)^2$,
we input $M_2$,
$M_3$,
$M_C$,
$\vartheta_{1}$,
$\vartheta_{2}$,
$\lambda_2$,
$\lambda_3$,
$\Re{\left( \lambda_6 \lambda_7^\ast \right)}$,
and $\Re{\left( \lambda_5^\ast \lambda_6 \lambda_7 \right)}$.
The angle $\vartheta_1$ is in either the first quadrant
or the fourth quadrant,
with a requirement that $c_1 \equiv \cos{\vartheta_1} > 0.9$
so that the $h_1 W^+ W^-$ coupling is within 10\% of its SM value.
The angle $\vartheta_2$ is in the first quadrant,
corresponding to a choice of the signs of the fields
of the new scalars $h_2$ and $h_3$.
%and the angle $\vartheta_3$ is free.
%The parameter $\lambda_2$ is positive.
%The parameter $\lambda_3 < M_C / v^2$.

We compute the oblique parameter $T$~\cite{musu, vikings}
and check that it is in its experimentally allowed domain~\cite{RPP}
$-0.04 < T < 0.20$. From eq.~\eqref{Mmatr} we then compute parameters 
$\lambda_1$,
$\lambda_4$,
$\lambda_5$,
$\lambda_6$, and
$\lambda_7$. 
Now we have all the parameters of the model
and we require them to satisfy both the unitarity conditions
and the BFB conditions.
Numerically diagonalizing the mass matrix~\eqref{Mmatr1},
we find the diagonalization matrix $R$
and we choose the overall sign of $R$ such that $R_{11} \equiv c_1 > 0$.
Having matrix elements $R_{11}$, $R_{12}$,
and $R_{13}$ we compute the cubic~\eqref{g3} and quartic~\eqref{g4}
Higgs couplings. 

In our numerical work we have found that for $c_1 \lesssim 0.99$,
the masses of the new scalar particles $ \left( \sqrt{M_C} \right.$,
$\sqrt{M_2}$,
and $\left. \sqrt{M_3} \right)$ of the 2HDM
can be no larger than $\sim 700$\,GeV.
For $c_1 \lesssim 0.95$,
they can be no larger than $\sim 550$\,GeV.
When $c_1$ becomes close to 1,
the masses of all three new scalars
may grow strongly to the order of TeV
and in that case they become almost identical.

\begin{figure}[t]
\begin{center}
\epsfig{file=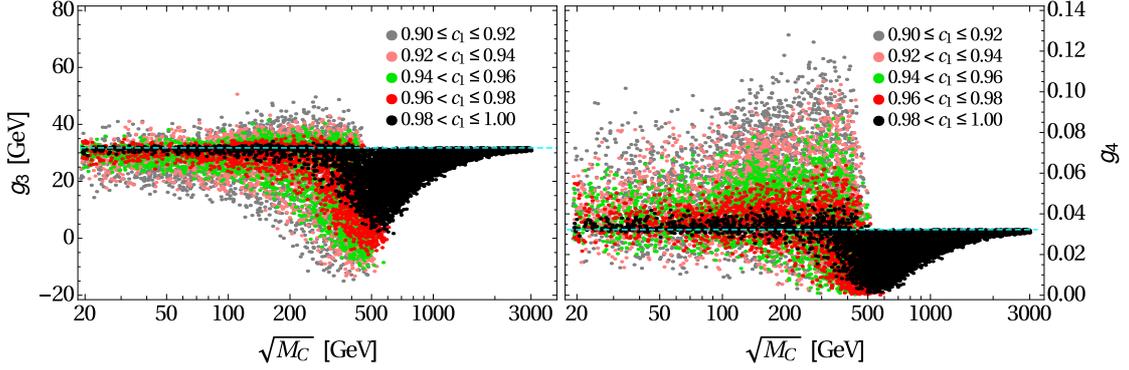,width=1.0\textwidth}
\vspace{-40pt}
\end{center}
\caption{The three-Higgs coupling $g_3$ (left panel) and of the four-Higgs coupling $g_4$ (right panel)\textit{versus}\/ $\sqrt{M_C}$ in the 2HDM for various values of $c_1$. The dashed lines mark the values of the couplings in the SM.}
\label{fig1}
\end{figure}

In figure~\ref{fig1} we have plotted the three- and four-Higgs couplings
$g_3$ and $g_4$ against the mass of the charged scalar $\sqrt{M_C}$. 
One sees that $g_3$ may be up to 1.6 times larger than in the SM,
but it may also be zero or even negative. On the other hand,
$g_4$ is always positive because of the boundedness from below of the potential
and may be up to four times larger than in the SM. 
For the masses of the new scalars up to 125~GeV,
$g_3$ has values in the range
$0.3 \lesssim g_3 \left/ g_3^\mathrm{SM} \right. \lesssim 1.6$
while $g_4$ has values in the range
$0 \lesssim g_4 \left/ g_4^\mathrm{SM} \right. \lesssim 3$.
For the masses up to 500~GeV the couplings reach their maximal values
but when the masses of the new scalars grow
beyond 1\,TeV the couplings $g_3$ and $g_4$ approach their SM values.

For a more detailed description of this model
and for a comparison of the couplings $g_3$ and $g_4$ in the 2HDM
with couplings calculated in other extensions of the SM
(Standart Model with the addition of one real singlet,
SM with two real singlets,
and 2HDM with the addition of one real singlet),
we suggest a look at ref.~\cite{musu}.  

\end{document}